\input epsf
\documentstyle{amsppt}
\overfullrule=0pt
\newcount\mgnf\newcount\tipi\newcount\tipoformule\newcount\greco
\tipi=2          
\tipoformule=0   

\global\newcount\numsec\global\newcount\numfor
\global\newcount\numapp\global\newcount\numcap
\global\newcount\numfig\global\newcount\numpag
\global\newcount\numnf
\global\newcount\numtheo

\def\SIA #1,#2,#3 {\senondefinito{#1#2}%
\expandafter\xdef\csname #1#2\endcsname{#3}\else
\write16{???? ma #1,#2 e' gia' stato definito !!!!} \fi}

\def \FU(#1)#2{\SIA fu,#1,#2 }

\def\etichetta(#1){(\veroparagrafo.\veraformula)%
\SIA e,#1,(\veroparagrafo.\veraformula) %
\global\advance\numfor by 1%
\write15{\string\FU (#1){\equ(#1)}}%
\write16{ EA #1 ==> \equ(#1)  }}

\def\etichettat(#1){\veroparagrafo.\veratheorema:%
\SIA e,#1,{\veroparagrafo.\veratheorema} %
\global\advance\numtheo by 1%
\write15{\string\FU (#1){\thu(#1)}}%
\write16{ TtH #1 ==> \thu(#1)  }}

\def\etichettaa(#1){(A\veraappendice.\veraformula)
 \SIA e,#1,(A\veraappendice.\veraformula)
 \global\advance\numfor by 1
 \write15{\string\FU (#1){\equ(#1)}}
 \write16{ EA #1 ==> \equ(#1) }}
\def\getichetta(#1){Fig. \verafigura
 \SIA g,#1,{\verafigura}
 \global\advance\numfig by 1
 \write15{\string\FU (#1){\graf(#1)}}
 \write16{ Fig. #1 ==> \graf(#1) }}
\def\retichetta(#1){\numpag=\pgn\SIA r,#1,{\verapagina}
 \write15{\string\FU (#1){\rif(#1)}}
 \write16{\rif(#1) ha symbol  #1  }}
\def\etichettan(#1){(n\verocapitolo.\veranformula)
 \SIA e,#1,(n\verocapitolo.\veranformula)
 \global\advance\numnf by 1
\write16{\equ(#1) <= #1  }}

\newdimen\gwidth
\gdef\profonditastruttura{\dp\strutbox}
\def\senondefinito#1{\expandafter\ifx\csname#1\endcsname\relax}
\def\BOZZA{
\def\alato(##1){
 {\vtop to \profonditastruttura{\baselineskip
 \profonditastruttura\vss
 \rlap{\kern-\hsize\kern-1.2truecm{$\scriptstyle##1$}}}}}
\def\galato(##1){ \gwidth=\hsize \divide\gwidth by 2
 {\vtop to \profonditastruttura{\baselineskip
 \profonditastruttura\vss
 \rlap{\kern-\gwidth\kern-1.2truecm{$\scriptstyle##1$}}}}}
\def\verapagina{
{\romannumeral\number\numcap}.\number\numsec.\number\numpag}}

\def\alato(#1){}
\def\galato(#1){}
\def\veroparagrafo{\number\numsec}\def\veraformula{\number\numfor}
\def\veraappendice{\number\numapp}
\def\verapagina{\number\pageno}\def\veranformula{\number\numnf}
\def\verafigura{{\romannumeral\number\numcap}.\number\numfig}
\def\verocapitolo{\number\numcap}\def\veranformula{\number\numnf}
\def\veratheorema{\number\numtheo}
\def\Eqn(#1){\eqno{\etichettan(#1)\alato(#1)}}
\def\eqn(#1){\etichettan(#1)\alato(#1)}
\def\TH(#1){{\etichettat(#1)\alato(#1)}}
\def\thv(#1){\senondefinito{fu#1}$\clubsuit$#1\else\csname fu#1\endcsname\fi}
\def\thu(#1){\senondefinito{e#1}\thv(#1)\else\csname e#1\endcsname\fi}

\def\Eq(#1){\eqno{\etichetta(#1)\alato(#1)}}
\def\eq(#1){\etichetta(#1)\alato(#1)}
\def\Eqa(#1){\eqno{\etichettaa(#1)\alato(#1)}}
\def\eqa(#1){\etichettaa(#1)\alato(#1)}
\def\dgraf(#1){\getichetta(#1)\galato(#1)}
\def\drif(#1){\retichetta(#1)}

\def\eqv(#1){\senondefinito{fu#1}$\clubsuit$#1\else\csname fu#1\endcsname\fi}
\def\equ(#1){\senondefinito{e#1}\eqv(#1)\else\csname e#1\endcsname\fi}
\def\graf(#1){\senondefinito{g#1}\eqv(#1)\else\csname g#1\endcsname\fi}
\def\rif(#1){\senondefinito{r#1}\eqv(#1)\else\csname r#1\endcsname\fi}
\def\bib[#1]{[#1]\numpag=\pgn
\write13{\string[#1],\verapagina}}

\def\include#1{
\openin13=#1.aux \ifeof13 \relax \else
\input #1.aux \closein13 \fi}

\openin14=\jobname.aux \ifeof14 \relax \else
\input \jobname.aux \closein14 \fi
\openout15=\jobname.aux
\openout13=\jobname.bib


\ifnum\tipoformule=1\let\Eq=\eqno\def\eq{}\let\Eqa=\eqno\def\eqa{}
\def\equ{}\fi


{\count255=\time\divide\count255 by 60 \xdef\hourmin{\number\count255}
        \multiply\count255 by-60\advance\count255 by\time
   \xdef\hourmin{\hourmin:\ifnum\count255<10 0\fi\the\count255}}

\def\oramin{\hourmin }

\def\data{\number\day/\ifcase\month\or january \or february \or march \or
april \or may \or june \or july \or august \or september
\or october \or november \or december \fi/\number\year;\ \oramin}

\def\titdate{ \ifcase\month\or January \or February \or March \or
April \or May \or June \or July \or August \or September
\or October \or November \or December \fi \number\day, \number\year;\ \oramin}

\setbox200\hbox{$\scriptscriptstyle \data $}

\newcount\pgn \pgn=1
\def\foglio{\number\numsec:\number\pgn
\global\advance\pgn by 1}
\def\foglioa{A\number\numsec:\number\pgn
\global\advance\pgn by 1}

\footline={\rlap{\hbox{\copy200}}\hss\tenrm\folio\hss}


\global\newcount\numpunt

\magnification=\magstephalf
\baselineskip=16pt
\parskip=8pt

\voffset=2.5truepc
\hoffset=0.5truepc
\hsize=6.1truein
\vsize=8.4truein 
{\headline={\ifodd\pageno\rightheadline \else \leftheadline \fi}}
\def\rightheadline{\it  {tralala}\hfil\tenrm\folio}
\def\leftheadline{\tenrm \folio \hfil\it  {Section $\ver$}}

\def\a{\alpha}
\def\b{\beta}
\def\d{\delta}

\def\f{\phi}
\def\g{\gamma}

\def\s{\sigma}

\def\G{\Gamma}

\def\S{\Sigma}

\def\1{{1\kern-.25em\roman{I}}}
\def\eu{{1\kern-.25em\roman{I}}}
\def\f1{{1\kern-.25em\roman{I}}}

\def\R{{\Bbb R}}  
\def\P{{\Bbb P}}  
\def\E{{\Bbb E}}  




\let\cal=\Cal

\def\DD{{\cal D}}
\def\EE{{\cal E}}

\def\MM{{\cal M}}

\def\PP{{\cal P}}

\def\SS{{\cal S}}

\def\chap #1#2{\line{\ch #1\hfill}\numsec=#2\numfor=1\numtheo=1}

\def\wh{\widehat}


\def\note#1{\footnote{#1}}

\def\frac#1#2{{#1\over #2}}

\def\text#1{\quad{\hbox{#1}}\quad}
\def\newpage{\vfill\eject}

\def\theo #1{\noindent{\thbf Theorem {#1} }}

\def\lemma #1{\noindent{\thbf Lemma {#1} }}

\def\endproof{$\diamondsuit$}
\def\remark{\noindent{\bf Remark: }}
\def\thanks{\noindent{\bf Acknowledgements: }}

\font\thbf=cmbxsl10 scaled\magstephalf

\font\ch=cmbx12
\font\ftn=cmr8

\font\it=cmti10
\font\bf=cmbx10

\newfam\msafam
\newfam\msbfam
\newfam\eufmfam
%
%
%
\def\hexnumber#1{%
  \ifcase#1 0\or 1\or 2\or 3\or 4\or 5\or 6\or 7\or 8\or
  9\or A\or B\or C\or D\or E\or F\fi}
\font\tenmsa=msam10
\font\sevenmsa=msam7
\font\fivemsa=msam5
\textfont\msafam=\tenmsa
\scriptfont\msafam=\sevenmsa
\scriptscriptfont\msafam=\fivemsa
\edef\msafamhexnumber{\hexnumber\msafam}%
%
%
\mathchardef\restriction"1\msafamhexnumber16
\mathchardef\ssim"0218
\mathchardef\square"0\msafamhexnumber03
\mathchardef\eqd"3\msafamhexnumber2C
\def\QED{\ifhmode\unskip\nobreak\fi\quad
  \ifmmode\square\else$\square$\fi}
\font\tenmsb=msbm10
\font\sevenmsb=msbm7
\font\fivemsb=msbm5
\textfont\msbfam=\tenmsb
\scriptfont\msbfam=\sevenmsb
\scriptscriptfont\msbfam=\fivemsb
\def\Bbb#1{\fam\msbfam\relax#1}
\font\teneufm=eufm10
\font\seveneufm=eufm7
\font\fiveeufm=eufm5
\textfont\eufmfam=\teneufm
\scriptfont\eufmfam=\seveneufm
\scriptscriptfont\eufmfam=\fiveeufm

\def\({\left(}
\def\){\right)}
%
%
%

%

{\headline={\ifodd\pageno\rightheadline \else \leftheadline \fi}}
\def\rightheadline{\it  {Mertens universality}\hfil\tenrm\folio}
\def\leftheadline{\tenrm \folio \hfil\it  {Section $\ver$}}

\font\tit=cmbx12
\font\aut=cmbx12
\font\aff=cmsl12
\overfullrule=0pt
\def\s{\char'31}
\centerline{\tit A tomography of  the GREM:  } \vskip.1truecm
\centerline{\tit beyond the REM conjecture \note{Research
supported in part by the DFG in the Dutch-German Bilateral
Research Group ``Mathematics of Random Spatial Models from Physics
and Biology'' and by the European Science Foundation in the
Programme RDSES.} } \vskip.2truecm
\vskip0.5truecm
\centerline{\aut Anton Bovier \note{ e-mail:
bovier\@wias-berlin.de}  } \vskip.1truecm \centerline{\aff
Weierstra\s {}--Institut} \centerline{\aff f\"ur Angewandte
Analysis und Stochastik} \centerline{\aff Mohrenstrasse 39, 10117
Berlin, Germany} \centerline{and} \centerline{\aff Institut f\"ur
Mathematik} \centerline{\aff Technische Universit\"at Berlin}
\centerline{\aff Strasse des 17. Juni 136,
12623 Berlin, Germany}
\vskip.4truecm
\centerline{\aut  Irina Kurkova\note{\ftn
e-mail: kourkova\@ccr.jussieu.fr}}
\vskip.1truecm
\centerline{\aff Laboratoire de Probabilit\'es et Mod\`eles
Al\'eatoires}
\centerline{\aff Universit\'e Paris 6}
\centerline{\aff 4, place Jussieu, B.C. 188}
\centerline{\aff 75252 Paris, Cedex 5, France}

\vskip0.2truecm\rm
\def\s{\sigma}
\noindent {\bf Abstract:} Recently, Bauke and Mertens conjectured
that the local statistics of energies in random spin systems with
discrete spin space should in most circumstances be the same as in
the random energy model. This was proven in a large class of
models for energies that do not grow too fast with the system
size. Considering the example of the generalized random energy
model, we show that the conjecture breaks down for energies
proportional to the volume of the system, and describe the far more complex
behavior that then sets in.

\noindent {\it Keywords: level statistics, random energy model,
generalized random energy model, extreme value theory, disordered
systems, spin glasses}

\noindent {\it AMS Subject  Classification: 60G70, 82B45}

{\headline={\ifodd\pageno\rightheadline \else \leftheadline \fi}}
\def\rightheadline{\it  {Beyond the REM conjecture}\hfil\tenrm\folio}
\def\leftheadline{\tenrm \folio \hfil\it  {Section $\ver$}}

\newpage

\chap{1. Introduction.}1 In a recent paper \cite{BaMe}, Bauke and
Mertens have formulated an interesting conjecture on the behavior
of local energy level statistics in disordered systems. Roughly
speaking, their conjecture can be formulated as follows. Consider
a random Hamiltonian, $H_N(\s)$, i.e.\ a random function from some
product space, $\SS^{N}$, where $\SS$ is a finite space, typically
$\{-1,1\}$,  to the real numbers.  We may assume for simplicity
that $\E H_N(\s)=0$. In such a situation, for typical $\s$,
$H_N(\s) \sim \sqrt N$, while $\sup_{\s}H_N(\s)\sim N$. Bauke and
Mertens then ask the following question: Given a fixed number,
$E$, what are the statistics of the values $N^{-1/2}H_N(\s)$ that
are closest to this number $E$, and how are configurations, $\s$,
for which these good approximants of $E$ are realized, distributed
on $\SS^{N}$? Their conjectured answer, which at first glance
seems rather surprising, is  simple: find $\delta_{N, E}$ such
that $\P( |N^{-1/2}H_N(\s)-E|\leq  b \delta_{N, E}) \sim
|\SS|^{-N} b$ for any constant $b>0$; then the collection of
points $ \delta_{N, E}^{-1} |N^{-1/2}H_N(\s)-E|$ over
     all  $\s\in \SS^{N}$
converges to a Poisson point process on $\R_+$, with
 intensity measure  the Lebesgue measure.
   Furthermore,  for any finite
$k$, the $k$-tuple of configurations
   $\s^1,\s^2,\dots, \s^k$, where the $k$ best approximations
are realized, is such that all of its elements have maximal
Hamming distance between each other. In other words, the
asymptotic behavior of these best approximants of $E$ is the same,
as if the random variables $H_N(\s)$ were all independent Gaussian
random
  variables with zero mean and variance $N$, i.e.\ as if we were dealing with the
random energy model (REM) \cite{Der1}; for this reason,  Bauke and
Mertens call this phenomenon ``universal REM like behavior''.

This conjecture was proven recently \cite{BK2} in a wide class of
models, including  mean field models and short range spin glass
models. In the case of Gaussian interactions, it was shown to hold
even for energies that diverge with the volume of the system, $N$,
as $E_N=c N^\a$, for $0\leq \a <\a_0$, where $\a_0$ is model
dependent.

Is is rather clear that the conjecture must break down in general for
$\a$ such that $cN^\a$ is of the order of the maximum of $H_N(\s)$. It
is a natural question to ask what will happen in this
regime. Naturally, the answers will become model dependent, and in
general very difficult to obtain. The only (non-trivial)
 models where we are able to carry out such an analysis
 in detail are the so-called {\it generalized random energy models}
   (GREMs) of
 Derrida \cite{Der2}. In these models, the extremal process was
 analyzed in full in \cite{BK1}. The result we obtain gives a somewhat
extreme microcanonical picture of the GREM, exhibiting in a somewhat
tomographic
way the distribution of states in a tiny vicinity of any value of the energy.

    Let us briefly recall the definition of the GREM.
We consider parameters $\a_0=1<\a_1,\dots, \a_n<2$ with
  $\prod_{i=1}^n \a_i=2$,  $a_0=0<a_1,\dots, a_n<1$,
   $\sum_{i=1}^n a_i=1$.  Let $\Sigma_N=\{-1, 1\}^N$
  be the space of $2^N$ spin configurations $\s$. Let
     $X_{\s_1\cdots \s_l}$,
   $l=1,\dots, n$, be
  independent  standard Gaussian random variables indexed by
  configurations $\s_1\dots\s_l\in \{-1,1\}^{N \ln(\a_1\cdots
    \a_l)/\ln 2}$.
  We define the Hamiltonian of the GREM as $H_N(\s)\equiv\sqrt{N} X_\s$,
with
$$
 X_{\s} \equiv\sqrt{a_1}X_{\s_1}+\cdots +\sqrt{a_n}X_{\s_1\cdots \s_n}.
\Eq(hmgrem)
$$
    Then $\hbox{cov}\, (X_{\s}, X_{\s'})=A(d_N(\s, \s'))$,
   where $d_N(\s, \s')=N^{-1}[\min \{i: \s_i\ne \s_i'\}-1]$,
  and $A(x)$ is a right-continuous
   step function  on $[0,1]$, such that, for
    any $i=0,1,\dots, n$,
    $A(x)=a_0+\cdots +a_i$, for
 $x\in [\ln(\a_0\a_1,\cdots\a_i)/\ln 2
    \,,\, \ln(\a_0\a_1,\cdots\a_{i+1})/\ln 2)$.

Set $J_0\equiv 0$, and, define, for $l>0$,
$$
J_l=\min\Big\{n\geq J>J_{l-1}: \frac{\ln(\a_{J_{l-1}+1}\cdots \a_J)}{
      a_{J_{l-1}+1}+\cdots +a_J}
  <\frac{\ln (\a_{J+1}\cdots \a_m)}{a_{J+1}+\cdots+a_m}
  \ \forall m\geq J+1\Big\}.
\Eq(ab.hmgrem.1)
$$
up to $J_k=n$.
Then, the  $k$  segments  connecting the points
    $(a_0+\cdots +a_{J_l}\,,\, \ln (\a_0\a_1\cdots \a_{J_l})/\ln
         2 )$, for  $l=0,1,\dots, k$
            form  the concave hull of the graph of the function $A(x)$.
 Let
$$
\bar a_l =a_{J_{l-1}+1}+a_{J_{l-1}+2}+\cdots +a_{J_l},\ \ \
  \bar \a_l=\a_{J_{l-1}+1}\a_{J_{l-1}+2}\cdots\a_{J_l}.
\Eq(aa).
$$
     Then
$$
 \frac{\ln \bar \a_1   }{\bar a_1  }< \frac{\ln \bar \a_2   }{\bar a_2  }<
  \cdots < \frac{\ln \bar \a_k   }{\bar a_k  }.
\Eq(mw)
$$
         Moreover, as it is shown in Proposition 1.4 of \cite{BK1},
      for any  $l=1,\dots, k$,  and for any
  $J_{l-1}+1\leq i<J_{l}$, we have
          $\ln (\a_{J_{l-1}+1} \cdots \a_i)/(a_{J_{l-1}+1}+
     \cdots +a_i) \geq \ln (\bar \a_{l})/\bar a_l$.
  Hence
$$
\frac{\ln \bar \a_l}{\bar a_l}= \min_{j=J_{l-1}+1, J_{l-1}+2,\dots, n}
     \frac{\ln (\a_{J_{l-1}+1}\dots \a_j)}{a_{J_{l-1}+1}+\cdots +a_j}.
\Eq(cru)
$$

      To formulate our results, we also need
 to recall from \cite{BK1} (Lemma 1.2)
  the point process of Poisson cascades $\PP^l$
 on $\R^l$.  It is   best understood in terms of the
  following iterative construction. If $l=1$,  $\PP^1$
      is the Poisson point process on $\R^1$ with
 the intensity measure $K_1 e^{-x}dx$.
 To construct $\PP^l$, we place the process
  $\PP^{l-1}$ on the plane  of the first $l-1$ coordinates and through
    each of its points draw a straight line orthogonal to this plane.
         Then we put on each
  of these lines independently a Poisson point process with intensity
 measure $K_l e^{-x}dx$. These points on $\R^l$
 form the process $\PP^l$.
  The  constants $K_1,\dots, K_l>0$
  (that are different from $1$ only in some degenerate cases)
 are defined in
         the formula (1.14) of \cite{BK1}.

      We will also need the following
  facts concerning $\PP^l$ from  Theorem 1.5 of \cite{BK1}.
           Let $\gamma_1>\gamma_2>\dots >\gamma_l>0$.
    There exists a constant $h>0$, such that,
  for all $y>0$,
$$
\P\big(\exists (x_1,\dots, x_l) \in \PP^l,
  \exists j=1,\dots, l: \gamma_1 x_1+\gamma_2x_2+\cdots +\gamma_j x_j>
         (\g_1+\cdots +\g_j)y\big)\leq  \exp(-h y ).
\Eq(fy)
$$
   Here and below we identify the measure $\PP^l$ with its support, when
suitable.
      Furthermore, for any $y \in \R$,
$$
\#\{(x_1,\dots, x_l) \in \PP^l : x_1\gamma_1+\cdots+ x_l\gamma_l>y\}<\infty
 \ \ a.s.
\Eq(fy1)
$$
 Moreover,  let $\beta>0$ be such that
$\beta \gamma_1>\cdots >\beta \gamma_l>1$.
 The integral
$$
\Lambda_l= \int\limits_{\R^l}
      e^{\beta (\gamma_1 x_1+\cdots \gamma_l x_l)}\PP^l(dx_1,\dots, dx_l).
 \Eq(yyi1)
 $$
 is understood as $\lim_{y \to -\infty }
        I_l(y)$ with
$$
\eqalign{
      I_l(y) & = \int\limits_{ (x_1,\dots, x_l) \in \R^l: \atop
 \exists i, 1\leq i \leq l: \g_1x_1+\cdots +\g_i x_i
   > (\g_1+\cdots +\g_i)y }
      e^{\beta(\gamma_1 x_1+\cdots +\gamma_l x_l)}
           \PP^l(dx_1,\dots, dx_l) \cr
&= \sum_{j=1}^l \int\limits_{ {(x_1,\dots, x_l) \in \R^l: \atop
   \forall i=1,\dots, j-1:    \g_1x_1+\cdots +\g_i x_i
   \leq  (\g_1+\cdots +\g_i)y }\atop
      \g_1x_1+\cdots +\g_j x_j
   >  (\g_1+\cdots +\g_j)y  }
      e^{\beta (\gamma_1 x_1+\cdots +\gamma_l x_l)}
           \PP^l(dx_1,\dots, dx_l,). }
 \Eq(sj)
$$
      It is finite, a.s., by Proposition 1.8 of \cite{BK1}.
   To keep the paper self-contained, let us recall how
  this fact can be established by induction starting from $l=1$.
  The integral \eqv(yyi1), in the case $l=1$,
 is understood as $\lim_{y\to -\infty} I_1(y)$.
  Here  $I_1(y)= \int\limits_{y}^{\infty} e^{\beta \g_1 x_1 } \PP_1(dx)$
   is finite, a.s., since $\PP_1$ contains  a finite number
   of points on $[y, \infty[$,  a.s.
   Furthermore, by \cite{BKL} or Proposition 1.8 of \cite{BK1},
       $\lim_{y\to-\infty}I_1(y)$ is finite, a.s.,
   since $\E\sup_{y'\leq y}(I_1(y')-I_1(y)) $
   converges to zero exponentially fast, as $y\to -\infty$,
  provided that $\beta \gamma_1>1$. If $l\geq 1$,
     each term in the representation \eqv(sj) is
determined and finite, a.s., by induction.
 In fact, to see this for the $j$th term, given any realization
 of $\PP^l$ in $\R^l$, take its projection on the plane
 of the first $j$ coordinates. Then by \eqv(fy1),
 there exists only a finite number of points
  $(x_1,\dots, x_j)$ of $\PP^j$,  such that
    $ \g_1x_1+\cdots +\g_j x_j
           >  (\g_1+\cdots +\g_j)y$, a.s.
  Whenever the first $j$ coordinates of a point of $\PP^l$
  in $\R^l$ are fixed, the remaining $l-j$ coordinates
  are distributed as $\PP^{l-j}$ on $\R^{l-j}$.
     Then the integral over the function
  $e^{\beta(\gamma_{j+1}x_{j+1}+\cdots +\g_l x_l)}$
      over these coordinates is defined by induction
  and is finite, a.s., provided that
  $\beta \gamma_{j+1}>\cdots >\beta \gamma_l>1$.
        Thus the $j$th term in \eqv(sj)
  is the sum of an a.s. finite number of terms
  and each of them is a.s. finite.
    Finally, again by Proposition 1.8 of \cite{BK1},
           $\lim_{y\to -\infty} I_l(y)$ is finite,
  a.s.,  since  $\E \sup_{y'\leq y}(I_l(y')-I_l(y))\to 0$
   as $y\to -\infty$ exponentially fast
provided that $\beta\gamma_1>\cdots >\beta \gamma_l>1$.

Let us define the constants $d_l$, $l=0,1,\ldots,k$, where
$d_0=0$ and
$$
d_l\equiv \sum_{i=1}^l\sqrt{\bar a_i 2\ln \bar \a_i}.
\Eq(ab.1)
$$
Finally, we define the domains $\DD_l$, for $l=0,\dots,k-1$, as
$$
\DD_l\equiv \left\{|y|<d_l+\sqrt {\frac {2\ln \bar\a_{l+1}}{\bar
a_{l+1}}}
\sum_{j=l+1}^k \bar a_j\right\}.
\Eq(ab.mwgrem.2)
$$
    It is not difficult to verify that $\DD_0 \subset \DD_1 \subset \cdots \subset  \DD_{k-1}$.
      We are now ready to formulate the main result of this paper.

\theo{\TH(thgr1)}{\it Let a sequence $c_N \in \R$
     be such  that $\lim\sup\limits_{N\to \infty} c_N \in \DD_0$ and
     $\lim\inf\limits_{N\to \infty}c_N \in \DD_0$.
           Then, the point process
$$
\MM_N^0=\sum_{\s \in \Sigma_N} \delta_{\big\{2^{N+1}
(2\pi)^{-1/2}e^{-c_N^2 N/2}
        \big|X_{\s} - c_N \sqrt{N }\big|\big\}}
\Eq(m0)
$$
 converges to the Poisson point process with  intensity measure
  the Lebesgue measure.

 \noindent   Let, for $l=1,\dots, k-1$,    $c \in \DD_l\setminus
   \overline{\DD_{l-1}}$ (where  $\overline{\DD_{l-1}}$
  is the closure of$\DD_{l-1}$).
    Define
$$
\tilde c_l =|c|- d_l,
\Eq(ccl)
$$
$$
\beta_l= \frac{\tilde c_l}{\bar a_{l+1}+\cdots +\bar a_k}, \ \ \ \
 \ \gamma_i=  \sqrt{ \bar a_i/ (2\ln \bar \a_i) }, \ \ i=1,\dots,l,
\Eq(not)
$$
and
$$
\eqalign{
   R_l(N)  = & \frac{2 (\bar \a_{l+1}\cdots \bar \a_k)^N
          \exp(-N\tilde c_l \beta_l/2) }{
       \sqrt{2\pi(\bar a_{l+1}+\cdots +\bar a_k)} }
      \prod\limits_{j=1}^l
 (4 N \pi \ln \bar \a_j)^{
       - \beta_l \gamma_j/2 } . }
\Eq(ml)
$$
  Then, the point process
$$
\MM_N^l = \sum_{\s \in \Sigma_N} \delta_{ \big\{ R_l(N)
       \big|\sqrt{a_1}X_{\s_1}+\cdots +\sqrt{a_n}X_{\s_1\dots \s_n}
    -  c\sqrt{N}\big| \big\} }
\Eq(pps)
$$
  converges to  mixed Poisson point process on $[0, \infty[$:
       given a realization of the random variable $\Lambda_l$,
 its intensity measure is $\Lambda_l dx$.
   The random variables $\Lambda_l$  is defined in terms of the Poisson
       cascades $\PP_l$ via
$$
\Lambda_l= \int\limits_{\R^l}
      e^{\beta_l(\gamma_1 x_1+\cdots \gamma_l x_l)}\PP^l(dx_1,\dots, dx_l).
 \Eq(yyi)
 $$
}
The next section will be devoted to the proof of this result.
Before doing this, we conclude the present section with a heuristic
interpretation of the main result.

Let us first look at \eqv(m0). This statement corresponds to the
REM-conjecture of Bauke and Mertens \cite{BaMe}.
It is quite remarkable
        that this conjecture holds in the case of the GREM
   for energies of the form  $c N $
   (namely for $c \in \DD_0$).

In the REM \cite{Der1}, $X_{\s}$ are $2^N$ {\it independent\/}
  standard Gaussian random variables and
 a
statement \eqv(m0)
   would hold for all $c$ with  $|c|< \sqrt{2\ln 2}$:
 it  is a well
known result from the theory of
independent random variables \cite{LLR}. The value
$c=\sqrt{2\ln 2}$ corresponds to
 the maximum of $2^N$ independent standard Gaussian
random variables, i.e., $\max_{\s \in \Sigma_N } N^{-1/2} X_{\s}
  \to \sqrt{2\ln 2}$ a.s.
Therefore, at the level $c=\sqrt{2\ln 2}$,
   one has the emergence of
the extremal process. More precisely, the point process
$$
\sum_{\s \in \S_N} \d_{\big\{\sqrt{2 N \ln 2 } \big(X_\s- \sqrt{2 N \ln 2 }+
    \ln (4\pi N \ln 2)/\sqrt{8 N \ln 2 } \big) \big\}},
\Eq(ab.grem.void)
$$
that is commonly written as
$\sum_{\s\in \S_N}\d_{u_N^{-1}(X_\s)}$
with
$$
u_N(x)= \sqrt {2N\ln 2} -\frac {\ln ( 4\pi N \ln 2) }{2\sqrt{2N\ln
2}}+\frac{x}{\sqrt{2N\ln 2}},
\Eq(ik.1)
$$
converges to the Poisson point process $\PP^1$ defined above
(see e.g. \cite{LLR}).  For $c>\sqrt {2\ln2}$,
the probability that one of the $X_\s$ will  be outside
of  the domain $\{|x|< c\sqrt{N}\}$, goes to
zero, and thus it makes no sense to consider such levels.

In the GREM,  $N^{-1/2}\max_{\s \in \Sigma_N }X_{\s}$
  converges to the value
     $d_k \in \partial D_{k-1}$ \eqv(ab.1)
(see Theorem 1.5 of \cite{BK1}) that is generally smaller
   than $\sqrt{2\ln 2}$.
Thus  it makes no sense to consider levels
 with $c \not\in \overline D_{k-1}$.
However, the REM-conjecture is not true for all levels
  in $\DD_{k-1}$,
but only in the smaller
 domain $\DD_0$.

To understand the statement of the theorem outside $\DD_0$,
we need to recall how the extremal process in the GREM is
related to the Poisson cascades introduced above. Let us set
$\S_{Nw_l}\equiv \{-1,1\}^{Nw_l}$ where
$$
w_l=\ln(\bar \a_1\cdots \bar \a_l)/\ln 2
\Eq(www)
$$
   with the notation \eqv(aa).   Let us also
  define the functions
$$
U_{l,N}(x)\equiv N^{1/2} d_l- N^{-1/2} \sum_{i=1}^l \g_i\ln( 4\pi N  \ln
\bar\a_i )/2+
N^{-1/2}x
\Eq(a.grem.2)
$$
with the notations \eqv(aa), \eqv(ab.1), \eqv(not),  and set
$$
\wh X^j_\s\equiv \sum_{i=1}^j\sqrt a_i X_{\s_1\dots\s_i}, \ \ \ \
\check X^j_\s
\equiv  \sum_{i=j+1}^{n}\sqrt a_i X_{\s_1\dots\s_i}.
\Eq(ab.grem.3.1)
$$
  From what was shown in \cite{BK1}, for any $l=1,\dots,k$,
    the point process,
$$
\EE_{l,N}\equiv
\sum_{\hat \s \in \S_{Nw_l}}
 \delta_{U^{-1}_{l,N}(\wh X^{J_l}_{\hat \s} )}
\Eq(ab.grem.4)
$$
converges in law to the Poisson cluster process, $\EE_l$,  given in terms
of the Poisson cascade, $\PP^l$, as
$$
\EE_l\equiv \int\limits_{\R^l} \PP^{(l)}(dx_1,\dots,dx_l)
\d_{\sum_{i=1}^l \g_ix_i}.
\Eq(ab.grem.5)
$$
In view of this observation, we can re-write the definition of the
process  $\MM_{N}^l$ as follows:
$$
\eqalign{
\MM_N^l = \sum_{\hat\s \in \Sigma_{w_lN}}
\sum_{\check\s\in \S_{(1-w_l)N}}
 \delta_{\big\{ R_l(N)
       \big|\check X^{J_l}_{\hat\s\check\s} - \sqrt{N}  \big[|c|-d_l
-N^{-1}(\G_{l, N}-U^{-1}_{l,N}(\wh X^{J_l}_{\hat \s})) \big] \big| \big\},
}}\Eq(ab.grem.6)
$$
with the abbreviation
$$
\G_{l, N}\equiv \sum_{i=1}^l\g_i \ln
  (4\pi N\ln\bar\a_i )/2 \Eq(gl)
$$
  ($c$ is replaced by $|c|$
  due to the symmetry of the standard Gaussian
  distribution).
The normalizing constant, $R_l(N)$, is chosen such that, for any
finite value, $U$, the point process
$$
\sum_{\check\s\in \S_{(1-w_l)N}}
 \delta_{\big\{ R_l(N)
       \big|\check X^{J_l}_{\hat\s\check\s} - \sqrt{N}  \big[|c|-d_l
-N^{-1}(\G_{l, N}-U) \big] \big| \big\},
}
\Eq(ab.grem.7)
$$
converges to the Poisson point processes on $\R_+$, with intensity
measure given by  $e^{U}$ times Lebesgue measure,
which is possible precisely because $c \in \DD_l \setminus
\overline{\DD_{l-1}}$,
  that is  $|c|-d_l$ is smaller that the a.s.\ limit of
   $N^{-1/2} \max_{\check \s\in \S_{(1-w_l)N}   }
   \check X^{J_l}_{\hat\s\check\s}$.
 This is
completely analogous to the analysis in the domain $\DD_0$.
Thus each term in the
sum over $\hat\s$ in \eqv(ab.grem.6) that gives rise to a ``finite''
 $U^{-1}_{l,N}(\wh X^l_{\hat \s})$, i.e., to an element of the extremal
 process of $\wh X^l_{\hat \s}$, gives rise to one Poisson process
 with
a random intensity measure in the limit of $\MM_{N}^l$. This explains
how the statement of the theorem can be understood,
and also  shows what the geometry of the configurations
realizing these mixed Poisson point processes will be.

  Let us  add that,
        if $c \in \partial \DD_{k-1}$,  i.e.
$|c|=d_{k}$, then one has the emergence
  of the extremal point process \eqv(ab.grem.4) with $l=k$, i.e.
$\sum_{\s \in \Sigma_N}
\delta_{\{\sqrt{N}(X_\s-d_k\sqrt{N}+N^{-1/2}\Gamma_{k, N}
)\}}$  converges   to \eqv(ab.grem.5) with $l=k$, see \cite{BK1}.


\bigskip
\chap{2. Proof of Theorem \thv(thgr1).}2

 Note that \eqv(yyi) is finite a.s.
  since $\gamma_1>\cdots >\gamma_l$ by \eqv(mw)
  and $\beta_l \gamma_l>1$ by the definition of $\beta_l$.
    Note also that $c$ can be replaced by $|c|$
  in \eqv(m0) and
   \eqv(pps)  due to the symmetry of the standard Gaussian
  distribution.

 Let $\MM_N^l(A)$ be the number
 of points of $\MM_N^l$ in a  Borel subset $A \subset \R_+$.
 We will show that for any finite disjoint union of intervals,
$A=\cup_{q=1}^p [a_q, b_q)$,
   the avoidance function converges
$$
\P(\MM_N^l(A)=0) \to \E \exp(- |A| \Lambda_l),
 \Eq(1d)
$$
  where of course $\Lambda_0=1$ in the case $l=0$.
Note that in that case, the right-hand side is the avoidance
function of a Poisson point process with intensity $1$, while in
all other cases, this is the avoidance function of a mixed Poisson
point process.

        To conclude the proof in the case $l=0$,
  it is enough to show that for any segment $A=[a, b)$
$$ \E \MM_N^0(A) \to (b-a), \ \ N \to \infty. \Eq(expe)$$
%
      Then the result \eqv(m0) would follow from Kallenberg's  theorem,
  see \cite{Ka} or  \cite{LLR}.

        In the cases $l=1,\dots, k-1$
   we will  prove   that the  family $\{\MM_N^l\}_{N=1}^\infty$
   is uniformly tight: by Proposition 9.1V of \cite{DV},
   this is equivalent to the fact that,  for any compact segment,
           $A=[a, b]$,  and  for any given  $\epsilon>0$,
 one can find a large enough integer,  $R$, such that
$$
\PP(\MM_N^l(A)>R)<\epsilon,\ \ \ \forall N\geq 1.
 \Eq(2d)
$$
          Finally, we will show that the limit
  of any weakly convergent subsequence of $\MM_N^l$
 is a simple point process, that is without double points
(see Definition 7.1IV in \cite{DV}).
       Theorem 7.3II of \cite{DV}  asserts that a simple point process
  is uniquely characterized by its avoidance function, which then
   implies the result \eqv(pps) claimed in Theorem \thv(thgr1).

   To  prove  \eqv(1d),  we need  the following lemma.

\lemma{\TH(l2)}{\it Let $A=\cup_{q=1}^p [a_q, b_q)$,
    $0\leq a_1<b_1<a_2<b_2<\cdots <a_q<b_q$, with
    $|A|=\sum_{q=1}^p(b_q-a_q)$.
Let $0<f<1$, $K(N)> 0$ be a polynomial in  $N$.
   We write  $K(N) f^N A\equiv \cup_{q=1}^p [K(N) f^N a_q, K(N) f^N b_q)$.

            For any $i=1,2,\dots, n$,
   any $\epsilon>0$, $\delta>0$ small enough,
   and  $M>0$,   there exists $N_0$, such that, for all
  $N\geq N_0$ and for all $y$, such that
$$
\eqalign{
\max\Big(& \max_{m=i+1,\dots, n}
    \frac{(a_{i}+\cdots +a_n)(2\ln \a_m+\cdots +2\ln \a_n+2\ln f+\epsilon )}
{a_m+\cdots +a_n},\cr
& (2\ln \a_{i+1}+\cdots +2\ln \a_n+2\ln f+\epsilon) \Big)  \leq y^2\leq M,
}
\Eq(hh1)
$$
 the probability,
 $$
 \P \Big(    \forall  \check\s\in \{-1,1\}^{N(\ln
 (\a_i\cdots\a_n)/\ln 2)} :
\Big|\frac{\check X^{i-1}_{\check\s}}{\sqrt{a_i+\cdots +a_n   }}
    - y \sqrt{N} \Big|
 \not \in  K(N) f^N A \Big),
\Eq(pit)
$$
with $\check X^{i-1}_{\check\s}$ defined by \eqv(ab.grem.3.1),
  is bounded from above and below, respectively, by
 $$
\exp\Big(-(1\pm \delta)|A|
          (2\pi)^{-1/2}  2 K(N) f^N \a_{i}^N \a_{i+1}^N \cdots \a_n^N
e^{-y^2 N /2} \Big).
\Eq(sh1)
$$
 }

\noindent{\it Proof.}
     Let us define the quantity
$$
P_N(i,y,f, K(N))
  \equiv \P\Big(   \exists  \check\s\in \{-1,1\}^{(\ln
   \a_{i+1}+\cdots \a_n)/\ln 2 }:   \Big|\frac{\check X^{i}_{\check\s}
  }{\sqrt{a_i+\cdots +a_n   }}
    - y \sqrt{N} \Big|
 \in   K(N) f^N A  \Big).
 \Eq(pi)
$$
        We will show that,  for any $\epsilon>0$ small enough and $M>0$
    large enough, we have
 $$
P_N(i,y,f, K(N))
  \sim  (2\pi)^{-1/2} 2 K(N) f^N |A| \a_{i+1}^N \cdots \a_n^N e^{-y^2
    N /2},
 \hbox{ as } \ N \to \infty,
\Eq(sh)
$$
   uniformly for the parameter $y$ in the domain
$$
\max_{m=i+1,\dots, n}
    \frac{(a_{i}+\cdots +a_n)(2\ln \a_m+\cdots +2\ln \a_n+2\ln f+\epsilon )}
{a_m+\cdots +a_n} \leq y^2\leq M.
\Eq(hh)
$$
   Then,   the probability \eqv(pit) equals
 $\big(1-P_N(i,y,f,K(N)) \big)^{\a_i^N}$, where the asymptotics
  of the quantity $P_N(i,y, f, K(N))$ is established in \eqv(sh).
Moreover, by the assumption
   \eqv(hh1),
$$
P_N(i,y,f, K(N))\leq
    (2\pi)^{-1/2} 2 K(N) |A|  \exp(-\epsilon N /2) \to 0.
\Eq(ab.hh1.1)
 $$
  Then the elementary inequality, $-x-x^2\leq  \ln(1-x) \leq -x$,
   that holds  for $|x|<1/2$, leads to \eqv(sh1).

  Therefore we concentrate  on the proof of
   the asymptotics  \eqv(sh).
   Let $X$ be a standard Gaussian random variable. Then
$$
P_N(n, y, f,K(N))=\P (|X-y\sqrt{N}| \in  K(N) f^N A )
  \sim  (2\pi)^{-1/2}  2 K(N) f^N |A| e^{-y^2 N /2}, \ \ N \to \infty,
\Eq(ko)
$$
       uniformly for $y^2\leq M $.  This implies
  \eqv(sh)  for $i=n$.
  Note also that
$$
P_N(i,y,f, K(N)) \leq \a_{i+1}^N \cdots \a_n^N \P(|X-y\sqrt{N}|
\in K(N) f^N A),
\Eq(ab.ko.1)
$$
  so that the upper bound for
         \eqv(sh) is immediate.
We will establish the lower bound    by induction
     downwards from $i=n$ to $i=1$,  using the identity
$$
\eqalign{
& P_N (i,y,f, K(N)) =
\int\limits_{-\infty}^{\infty}
 \frac{dt \,e^{-t^2/2}}{\sqrt{2\pi}}
    \Biggl(1-\Big[1-\cr
&\quad\quad -P_N\Big(i+1, \frac{ \sqrt{a_i+\cdots +a_n}y \sqrt{N}
    -\sqrt{a_i}t}{
  \sqrt{N(a_{i+1}+\cdots +a_n)}}, f, \frac{ \sqrt{a_i+\cdots +a_n}  }{
  \sqrt{a_{i+1}   +\cdots +a_n} }K(N)
\Big) \Big]^{\alpha_{i+1}^N}\Biggr).
}
 \Eq(id)
$$
      By the induction hypothesis for $i+1$,
 $$
\eqalign{
P_N & \Big(i+1, \frac{ \sqrt{a_i+\cdots +a_n}y-\sqrt{a_i}t}{
     \sqrt{N(a_{i+1}+\cdots +a_n)}}, f,
 \frac{ \sqrt{a_i+\cdots +a_n}  }{ \sqrt{a_{i+1}
  +\cdots +a_n} } K(N)\Big) \cr
& \sim
 (2\pi)^{-1/2}
  \frac{ \sqrt{a_i+\cdots +a_n}  }{ \sqrt{a_{i+1}
  +\cdots +a_n} }
 2 K(N) f^N |A| \a_{i+2}^N
 \a_{i+3}^N \cdots \a_n^N e^{-
   \frac{(\sqrt{a_i+\cdots +a_n}y \sqrt{N}-\sqrt{a_i}t)^2}{
     2(a_{i+1}+\cdots +a_n)} },
 }
\Eq(kl)
$$
  uniformly for all $y,t$ that satisfying
$$
 \eqalign{  \max_{m=i+2,\dots, n} &
    \frac{(a_{i+1}+\cdots +a_n)(2\ln \a_m+\cdots +2\ln \a_n+2\ln f+\epsilon_{i+1} )}
 {a_m+\cdots +a_n} \cr
   & \leq  \Big( \frac{ \sqrt{a_i+\cdots +a_n}y \sqrt{N} -\sqrt{a_i}t}{
  \sqrt{N(a_{i+1}+\cdots +a_n)}}\Big)^2 \leq M_{i+1},
 }
\Eq(uu)
$$
   for any $\epsilon_{i+1}>0$ small enough and $M_{i+1}>0$ large enough.
    The right-hand side of this inequality reads
$$
\eqalign{
\sqrt{N} T^{-}_1(y) & \equiv \sqrt{N}\frac { \sqrt{a_i+\cdots +a_n}y
   -\sqrt{ a_{i+1}+\cdots +a_n}M_{i+1} } {\sqrt{a_i}}  \leq t \cr
    & \leq \sqrt{N}
          \frac { \sqrt{a_i+\cdots +a_n}y
   +\sqrt{ a_{i+1}+\cdots +a_n}M_{i+1}}
 {\sqrt{a_i}} =\sqrt{N}T^{+}_1(y).
}
\Eq(ab.kl.1)
$$
   Obviously, the left-hand side of \eqv(uu) holds
     for all $t \in (-\infty, \infty)$, if
        $\ln \a_n+\cdots +\ln \a_{i+2} +2\ln f<0$
and $\epsilon_{i+1}$ is small enough.
   Otherwise,  it holds, if either
$$
\eqalign
{t  & \geq \frac{\sqrt{N} }{\sqrt{a_i} }
  \max\limits_{m =i+2,\dots, n :
     \atop \ln \a_n+\cdots +\ln \a_m +2\ln f \geq 0 } \Big(
   \sqrt{a_i+\cdots +a_n} y \cr
& \ \ \ \ \ {}+ \frac{a_{i+1}+\cdots +a_n}{\sqrt{
   a_m+\cdots +a_n}}
   \sqrt{2\ln \a_m+\cdots +2\ln \a_n+2\ln f+\epsilon_{i+1}}\Big)\
  \equiv \sqrt{N}T^+_2(y), }
\Eq(uu1)
$$
or
$$
\eqalign{
t & \leq \frac{\sqrt{N} }{ \sqrt{a_i} }
\min\limits_{m =i+2,\dots, n :
     \atop \ln \a_n+\cdots +\ln \a_m +2\ln f \geq 0 } \Big(
   \sqrt{a_i+\cdots +a_n} y\cr
 & \ \ \ \ \ {}- \frac{a_{i+1}+\cdots +a_n}{\sqrt{a_m
    +\cdots +a_n}}
   \sqrt{2\ln \a_m+\cdots +2\ln \a_n+2\ln f+\epsilon_{i+1}}\Big)
 \equiv \sqrt{N}T^-_2(y).
}
\Eq(uu2)
$$
Let us put for convenience $T^+_2(y)=-\infty$ and
              $T^-_{2}(y)=\infty$,  if
  $2\ln \a_n+\cdots +2\ln \a_{i+2} +2\ln f<0$.
  Finally,
$$
\a_{i+1}^N P_N\Big(i+1, \frac{ \sqrt{a_i+\cdots +a_n}y-\sqrt{a_i}t}{
     \sqrt{N(a_{i+1}+\cdots +a_n)}}, f,  \frac{ \sqrt{a_i+\cdots +a_n}  }
{ \sqrt{a_{i+1}
  +\cdots +a_n} } K \Big) \to 0,
\Eq(gg)
$$
   uniformly in the domain where
  $$
\Big( \frac{ \sqrt{a_i+\cdots +a_n}y \sqrt{N} -\sqrt{a_i}t}{
  \sqrt{N(a_{i+1}+\cdots +a_n)} }\Big)^2 \geq 2\ln \a_{i+1}+\cdots
 +2\ln \a_n+2\ln f+\epsilon_{i+1}.
\Eq(do)
$$
 This domain is equivalent to
  $-\infty <t< +\infty$,
   if  $2\ln \a_{n}+\cdots +2\ln \a_{i+1} +2\ln f<0$
  and $\epsilon_{i+1}>0$ is small enough.
   Otherwise,
 it is reduced  to the union of the domains
 $$
\eqalign{
t &\geq \frac{\sqrt{N}}{\sqrt{a_i} }
 \Big(\sqrt{a_i+\cdots a_n}y +\sqrt{(a_{i+1}+\cdots + a_n)
    (2\ln \a_{i+1}+\cdots +2\ln \a_n +2\ln f+\epsilon_{i+1}) } \Big) \cr
 & \equiv T^{+}_3(y) \sqrt{N} }
\Eq(ab.do.1)
$$
and
$$
\eqalign{ t & \leq \frac{\sqrt{N}}{\sqrt{a_i} }
 \Big( \sqrt{a_i+\cdots a_n}y - \sqrt{(a_{i+1}+\cdots + a_n)
    (2\ln \a_{i+1}+\cdots +2\ln \a_n +2\ln f+\epsilon_{i+1}) } \Big)\cr
  &  \equiv T^{+}_3(y) \sqrt{N} }.
\Eq(ab.do.2)
$$
  Then, using the elementary inequalities
$$
-x-x^2\leq  \ln(1-x) \leq -x,
\ \ \ 1+x\leq e^x \leq 1+x+x^2 \ \ \hbox{for }|x|<1/2,
\Eq(elem)
$$
         it is easy to deduce from \eqv(id), \eqv(kl), and \eqv(gg)
   the following   asymptotic lower bound, if
   $2\ln \a_{n}+\cdots +2\ln \a_{i+1}+2\ln f \geq 0$:
$$
 \eqalign{
& P(i,y, f, K(N))  \geq
      (2\pi)^{-1}
  \frac{ \sqrt{a_i+\cdots +a_n}  }{ \sqrt{a_{i+1}
  +\cdots +a_n} }
 2 K(N) f^N \alpha_{i+1}^N  \a_{i+2}^N
 \a_{i+3}^N \cdots \a_n^N \cr
 & {}\times \Big(\int\limits_{T^-_{1}(y)\sqrt{N} }^{
    \min (T^-_2(y), T^-_3(y))\sqrt{N}}+
    \int\limits_{\max (T^+_2(y), T^+_3(y))\sqrt{N} }^{T^+_1(y)\sqrt{N}} \Big)
 e^{- \frac{(\sqrt{a_i+\cdots +a_n}y \sqrt{N}-\sqrt{a_i}t)^2}
   {2(a_{i+1}+\cdots +a_n) } }
   e^{-t^2/2} dt.
}
\Eq(bon)
$$
  If $2 \ln \a_{i+1}+\cdots +2\ln \a_n+2\ln f < 0$,
 then from the same assertions we deduce
  the same bound, but with the domain of integration ranging over
 the entire interval $[T^-_1(y)\sqrt{N}, T^+_1(y)\sqrt{N}]$.
 By the change of variables,
$$
s= \frac{\sqrt{a_i+\cdots +a_n} t-\sqrt{a_i}y}{\sqrt{a_{i+1}+\cdots +a_n}},
 \Eq(chv)
$$
  the right-hand side of  \eqv(bon) equals
$$
 \frac{ 2 K(N)}{2\pi} f^N \alpha_{i+1}^N  \a_{i+2}^N
 \a_{i+3}^N \cdots \a_n^N e^{-y^2 N/2}
         \Big(\int\limits_{S^-_{1}(y)\sqrt{N} }^{
    \min (S^-_2(y), S^-_3(y))\sqrt{N}}+
    \int\limits_{\max (S^+_2(y), S^+_3(y))\sqrt{N} }^{S^+_1(y)\sqrt{N}} \Big)
   e^{-s^2/2}ds
\Eq(fl)
 $$
 where
$$
 S_1^{-}(y), S_1^{+}(y)= \frac{\sqrt{a_{i+1}+\cdots+a_n}y \pm \sqrt{a_i+\cdots+a_n}M_{i+1}}
   {\sqrt{a_i}},
\Eq(s1)
$$
$$
\eqalign{
S_2^-(y) = \min\limits_{m =i+1,\dots, n :
     \atop \ln \a_n+\cdots +\ln \a_l +2\ln f \geq 0 }
  & \sqrt{(a_{i+1}+\cdots+a_n)/a_i} \cr
   &  {} \times \Big( y- \frac{\sqrt{a_i+\cdots+a_n }}{\sqrt{ a_m+\cdots+a_n} }
     \sqrt{\ln \a_m+\cdots+\ln \a_n+\ln f+\epsilon_{i+1}}  \Big),
 }
\Eq(t1-)
$$
$$
\eqalign{
S_2^+(y) = \max\limits_{m =i+1,\dots, n :
     \atop \ln \a_n+\cdots +\ln \a_m +2\ln f \geq 0 }
  & \sqrt{(a_{i+1}+\cdots+a_n)/a_i} \cr
   &  {} \times \Big( y+\frac{\sqrt{a_i+\cdots+a_n }}{\sqrt{ a_m
  +\cdots+a_n} }
     \sqrt{\ln \a_l+\cdots+\ln \a_n+\ln f+\epsilon_{i+1}}  \Big),
 }
\Eq(t1+)
$$
  if $T^{\pm }_2(y)$ are finite,
    and, of course, $S^+_2(y)=-\infty$, if $T^+_2(y)=-\infty$,
   $S^-_2(y)=+\infty$, if  $T^-_2(y)=+\infty$, and finally
$$
S_3^\pm(y) =
    \frac{\sqrt{a_{i+1}+\cdots +a_n}y \pm
    \sqrt{a_i+\cdots +a_n}\sqrt{\ln \a_{i+1}+\cdots+\ln \a_n+
   \ln f +\epsilon_{i+1} }  }{\sqrt{a_i}}.
\Eq(ab.t1++)
$$

       Now let us take  any $\epsilon>\epsilon_{i+1}$ and $M=M_{i+1}$
   Then, there exist $\delta>0$ and $Q>0$, such that,
        for all $y\geq 0$ satisfying \eqv(hh), we have
   $S^{-}_1(y)\leq - Q $ and  $\min (S^-_2(y), S^-_3(y)) \geq \delta $;
      and  for all  $y<0$ satisfying \eqv(hh), we have
   $S^{+}_1(y)\geq Q $ and  $\max (S^+_2(y), S^+_3(y)) \leq -\delta $.
    Hence
 $$
(2\pi)^{-1/2} \Big( \int\limits_{S^-_{1}(y)\sqrt{N} }^{
         \min (S^-_2(y), S^-_3(y))\sqrt{N} }+
      \int\limits_{\max (S^-_2(y), S^-_3(y))\sqrt{N} }^{S^+_1(y)\sqrt{N}} \Big)
  e^{-s^2/2}ds \geq  (2\pi)^{-1/2} \int_{-Q \sqrt{N} }^{\delta
    \sqrt{N}} e^{-s^2/2}ds \to 1,
 \Eq(iil)
$$
   as $N\to \infty$.  In the case when
   $2\ln_{n}+\cdots +2\ln \a_{i+1}+2\ln f<0$, we have
    the analogue of \eqv(bon) with the integral over $[T^-_1(y)\sqrt{N},
         T^+_1(y)\sqrt{N}]$, and
    by the same change we get the bound
 $$
(2\pi)^{-1/2} \int\limits_{S^-_{1}(y)\sqrt{N}}^{S^+_1(y)\sqrt{N}}
  e^{-s^2/2}ds \geq  (2\pi)^{-1/2}
          \int\limits_{-Q \sqrt{N} }^{ Q \sqrt{N}} e^{-s^2/2}ds \to 1,
   \  \  N \to \infty.
 \Eq(iim)
$$
             Since  $\epsilon_{i+1}$ [resp.\ $M_{i+1}$]  could be chosen
  arbitrarily small [resp.\ large], by the induction hypothesis,
    the estimates \eqv(bon),
    \eqv(fl), and \eqv(iil), \eqv(iim)
   show that, for any $\epsilon>0$ small enough,
   and $M>0$ large enough, the assertion \eqv(sh) holds
    uniformly in the domain
  \eqv(hh). This finishes the proof of the lemma.  \endproof

Lemma \thv(l2) implies the next lemma.

\lemma{\TH(lem2)}{\it
 Let  $l\in \{0,\dots, k-1\}$,  $c$  be  with
  $|c|<\sqrt{2\ln \bar \a_{l+1}  (\bar a_{l+1}
+\cdots + \bar a_k)/\bar a_{l+1} }$. For
  any $\epsilon, \delta>0$ small enough, and
    $M>0$,  there exists  $N_0=N_0(\epsilon, \delta, M)$,
   such that,
   for all $N\geq N_0$,
 the probability
 $$
\eqalign{
 \P \Big(  &  \forall  \check\s\in\{-1,1\}^{(1-w_{l})N}:
 \Big|\frac{\check X^{J_{l}}_{\check\s} }
   {\sqrt{\bar a_{l+1}+\cdots +\bar a_k   }}
    - (|c|+z)\sqrt{N} \Big|
 \not\in  K(N) e^{c^2 N/2} (\bar \a_{l+1} \cdots \bar \a_k)^{-N} A \Big)
}
 \Eq(pit1)
$$
    is bounded from above and below, respectively, by
 $$
\exp\Big(-(1\pm \delta)
       (2\pi)^{-1/2} 2 K(N) |A| e^{-(2|c|z+z^2) N /2} \Big)
    \Eq(sh2n)
 $$
   for any $-\epsilon<z<M$.}

\noindent{\it Proof.}
   If   $|c|<\sqrt{2\ln \bar \a_{l+1}  (\bar a_{l+1}
   +\cdots + \bar
 a_k)/\bar a_{l+1} }$,  then
 by \eqv(cru)  we have
   $e^{c^2/2}(\bar \a_{l+1}\cdots \bar \a_k)^{-1}<1$
  and with some  $\epsilon_0>0$ small enough:
$$
\eqalign{
\max\Big( & \max_{m=J_{l}+2,\dots, n}\!\!\!\!\!\!\!\!
    \frac{(a_{J_{l}+1}+\cdots +a_n)(2\ln \a_m+\cdots +2\ln \a_n+2 ( c^2/2 -\ln (\bar
   \a_{J_{l+1}}\cdots \bar \a_{J_k}) ) +\epsilon_0 )}
{a_m+\cdots +a_n}, \cr
 & (2\ln \a_{J_{l}+2}+\cdots +2\ln \a_n+2
  ( c^2/2 -\ln (\bar
   \a_{J_{l+1}}\cdots \bar \a_{J_k}))
+\epsilon_0 \Big)   < c^2. }
\Eq(hh1.1)
$$
     This last inequality
  remains true with $c^2$ replaced in the left-hand side
     by $(|c|+z)^2$
  if $z>-\epsilon$ with $\epsilon>0$ small enough.
 Then  Lemma~\thv(l2) applies
  with $i=J_{l}+1$ and $f= e^{c^2 /2} (\bar \a_{l+1} \cdots \bar \a_k)^{-1}$
 and gives the asymptotics \eqv(sh2n).\endproof

        Lemma \thv(lem2)  with  $l=0$,
  $z=0$,  $K(N)= \sqrt{2\pi}/2$
 implies immediately the convergence
 of the avoidance function \eqv(1d) in the case $l=0$.
         To conclude the proof  of \eqv(m0),
   let us note that
$$
\E\MM_N^0(A) = \sum_{\s \in \Sigma_N}
    \P\big(|X_\s-c_N \sqrt{N}| \in 2^{-N-1} (2\pi) e^{c_N^2 N/2}A \big)
\Eq(eeee)
$$
 is the sum of $2^N$ identical terms, each of them
   being $2^{-N}|A|(1+o(1))$  by the trivial estimate
     for standard Gaussian random variables \eqv(ko).
Then \eqv(eeee) converges to $|A|$ and the proof
  of \eqv(m0) is finished.

      To prove the convergence of  the avoidance function  \eqv(1d)
  in the case $l\geq 1$, let us write the event
$\{\MM_N^l(A)=0\}$
  in terms of  the functions $U_{l,N}$  defined in \eqv(a.grem.2) as
$$
\eqalign{
& \{\MM_N^l(A)=0\}\cr
& =  \Big\{\forall \hat \s \in \Sigma_{w_l N},
     \check \s \in \Sigma_{(1-w_l)N}: \big|\check X^{J_l}_{\hat \s
       \check \s}
     -\sqrt N\big[\tilde c_l  + N^{-1}\big(\Gamma_{l, N} -
   U_{l,N}^{-1}(\wh X^{J_l}_{\hat \s})\big)\big]\big|
        \not\in R_l(N)^{-1} A \Big\}
}
\Eq(proby)
$$
  with the abbreviations \eqv(www), \eqv(ab.grem.3.1), \eqv(gl).
      Let us introduce the  following event with
  a parameter $y>0$:
$$
\eqalign{
B^l_N(y)=\Big\{ & \forall j=1,\dots, l,\forall\hat\s\in \S_{w_l N}:\cr
& 2\Gamma_{j, N } - 2 N d_j - (\gamma_1+\cdots +\gamma_j)y
         <U_{j,N}^{-1}(\wh X^{J_j}_{\hat\s})
    <y(\gamma_1+\cdots +\gamma_j) \Big\}.
   }
\Eq(bbb)
$$
  By the convergence
   \eqv(ab.grem.4) to \eqv(ab.grem.5),
   the property  \eqv(fy) and the symmetry of the standard
  Gaussian distribution, the probability of the complementary event
  satisfies the following  bound:
 $$
\lim\sup_{N\to \infty} \P(\bar B_N^l(y)) \leq 2\exp(- h y),
\Eq(bb)
$$
with some constant $h>0$.
      Now, let us fix any arbitrarily large $y>0$ and consider
  $$\eqalign{
\P(\MM_N^l(A)=0) = & \E\big[ \1_{\{B_N^l(y)\}}
    \E(\1_{\{\MM_N^l(A)=0 \}}\mid \wh X^{J_j}_{\hat \s},
  \forall_{j=1}^l, \forall \wh \s \in \Sigma_{w_l N} )\big]\cr
 & {} + \E\big[ \1_{\{\bar B_N^l(y)\}}
\E(\1_{\{\MM_N^l(A)=0 \}}\mid \wh X^{J_j}_{\hat \s},
  \forall_{j=1}^l,
  \forall \wh \s \in \Sigma_{w_l N} )\big]. }
\Eq(toto)
$$
      Due to the representation
       \eqv(proby), the conditional  expectation
      $ \E(\1_{\{\MM_N^l(A)=0 \}}\mid \wh X^{J_j}_{\hat \s},
  \forall_{j=1}^l,
 \forall \wh \s \in \Sigma_{w_l N}
)$  can
       be viewed as the product over $\wh \s \in \Sigma_{w_l N}$
   of the quantities \eqv(pit1) with
$$
|c|=
\frac{\tilde c_l }{\sqrt{\bar a_{l+1}+\cdots +\bar a_k}},\ \ \ \
 K(N)= \frac{\sqrt{2\pi}}{2}
   \prod\limits_{j=1}^l  (4 N \pi \ln \bar \a_j)^{
       \beta_l \gamma_j/2 },
\Eq(cc)
$$
  and
$$
z= z(\hat \s) = (\bar a_{l+1}+\cdots +\bar a_k)^{-1/2}
           N^{-1}\Big(\Gamma_{l, N} -
   U_{l,N}^{-1}(\wh X^{J_l}_{\hat \s})\Big), \ \ \hat \s\in \Sigma_{w_l N}.
\Eq(zz)
$$
       Furthermore,  on $B_N^l(y)$,  we have
   $z(\wh \s) \in
     (-\epsilon \,,\, \frac{2 d_l}{
 \sqrt{\bar a_{l+1}+\cdots + \bar a_k }}
      +\epsilon)$  $\forall   \hat \s\in \Sigma_{w_l N}$
  (with some small enough $\epsilon>0$),
           so that Lemma \thv(lem2) applies
   to  $\1_{\{B_N^l(y)\}}
    \E(\1_{\{\MM_N^l(A)=0 \}}\mid \wh X^{J_j}_{\hat \s},
  \forall_{j=1}^l, \forall \wh \s \in \Sigma_{w_l N} )$.
  Hence,  by \eqv(toto) and
 by Lemma \thv(lem2),  for any
 $\delta>0$ small enough, there exists $N_0(\delta, y)$ such that
  for all $N\geq N_0$
$$
\eqalign{
& \E \Big[\prod\limits_{\hat \s\in \Sigma_{w_l N} }
\exp \Big(-(1-\delta) (2\pi)^{-1/2} 2 K(N)|A|
      e^{-  \big(2|c| z(\hat \s )+
          z^2(\hat \s ) \big )N/2 }
  \Big)\Big] + \P(\bar B_N^l(y))\cr
&{} \geq
\E \Big[ \1_{\{ B_N^l(y)\}}
\prod\limits_{\hat \s\in \Sigma_{w_l N} }
\exp \Big(-(1-\delta) (2\pi)^{-1/2} 2 K(N)|A|
      e^{-  \big(2|c| z(\hat \s )+
            z^2(\hat \s ) \big )N/2 }
  \Big)\Big] + \P(\bar B_N^l(y))\cr
 & {}\geq  \P(\MM_N^l(A)=0)  \cr
 &{} \geq \E \Big[ \1_{\{ B_N^l(y)\}}
\prod\limits_{ \hat \s\in \Sigma_{w_l N}}
\exp \Big(-(1+\delta) (2\pi)^{-1/2} 2 K(N)|A|
      e^{-\big(2|c| z(\hat \s )+
          z^2(\hat \s) \big )N/2 }
  \Big)  \Big]\cr
&\geq \E \Big[\prod\limits_{ \hat \s\in \Sigma_{w_l N}}
\exp \Big(-(1+\delta) (2\pi)^{-1/2} 2 K(N)|A|
      e^{-\big(2|c| z(\hat \s )+
          z^2(\hat \s) \big )N/2 }
  \Big)  \Big]
-\P(\bar B_N^l(y)).
}
 \Eq(gpp)
$$
      Using the convergence \eqv(ab.grem.4) to
   \eqv(ab.grem.5),  we  derive that for any
   $y>0$ large  enough and $\delta>0$ small enough
$$
\eqalign{
& \E \prod_{ (x_1,\dots, x_l) \in \PP_l}
    \exp(-(1 - \delta) |A| e^{\beta_l(\gamma_1 x_1+\cdots +\gamma_l x_l)})
         + \lim\sup_{N\to \infty} \P(\bar B_N^l(y))\cr
 & {}\geq  \lim \sup_{N \to \infty}\P(\MM_N(A)=0)
   \geq  \lim \inf_{N \to \infty}\P(\MM_N(A)=0) \cr
  & {} \geq  \E \prod_{ (x_1,\dots, x_l) \in \PP_l}
    \exp(-(1 +\delta) |A| e^{\beta_l(\gamma_1 x_1+\cdots +\gamma_l x_l)})
        -\lim\sup_{N\to \infty} \P(\bar B_N^l(y)).
}
\Eq(lal)
$$
 Thus
    \eqv(lal) and  \eqv(bb) imply the following bounds:
 $$
\eqalign{
& \E \exp(-(1-\delta)|A|\Lambda_l)
    +2\exp(-h y) \geq \lim \sup_{N \to \infty} \P(\MM_N^l(A)=0)  \cr
       &{} \geq   \lim \inf_{N \to \infty} \P(\MM_N^l(A)=0)
          \geq \E \exp(-(1+\delta)|A|\Lambda_l))
  -2\exp(-hy).
}
\Eq(aja)
$$
      Since $y>0$  can be chosen arbitrarily large
  and $\delta>0$ fixed arbitrarily small, this finishes
  the proof of the convergence of the avoidance function
    \eqv(1d) in the case of $l=1,2,\dots,
  k-1$.

   To proceed with  the proof of
   tightness \eqv(2d), we need the following lemma.

\lemma{\TH(lem3)}{\it
      Let   $l\in \{0,\dots, k-1\}$,
 $|c|<\sqrt{2\ln \bar \a_{l+1}  (\bar a_{l+1}
+\cdots + \bar a_k)/\bar a_{l+1} }$,
     $K(N)>0$ is polynomial  in $N$, $z \in \R$.
 For any
  segment $B \subset \R_+$,  let us define
an integer-valued random variable
$$ \eqalign{
&T^{c,z,K(N)}_{l, N}(B) \cr
&    =\#\Big\{\check\s\in\Sigma_{(1-w_{l})N}:
 \Big|\frac{\check X^{J_{l}}_{\check\s} }
   {\sqrt{\bar a_{l+1}+\cdots +\bar a_k   }}
    - \sqrt{N}(|c|+z) \Big|
 \in  K(N) e^{c^2 N/2} (\bar \a_{l+1} \cdots \bar \a_k)^{-N}B \Big\}.}
\Eq(TT)
$$
(i)   For any bounded segment $A \subset \R_+$,  any
         $\epsilon, \delta >0$ small enough and
   $M>0$ there exists $N_0=N_0(\delta, M, \epsilon)$
            such that for all $N\geq N_0$,
  for any $B\subset A$ and any $z \in ]-\epsilon, M[$
   we have:
$$
\P\big(T^{c,z,K(N)}_{l, N}(B) \geq 1\big)
  \leq (1+\delta) |B| K(N) (2/\sqrt{2\pi})e^{-(2|c|z+z^2)N/2}.
\Eq(l3i)
$$

\noindent (ii)  For any  bounded segment $A \subset \R_+$, any
    $\delta >0$ small enough, $K>0$ large enough and
     $M>0$ there exists $N_0=N_0(\delta, M, K)$
           such that for all $N\geq N_0$,
  for any segment $B\subset A$ with $|B|<K^{-1}$  and for any
$$
   z=z_N\in \Big] \frac{\ln( 2 K(N)/\sqrt{2\pi}) -\ln K }{|c| N}
     \,\,,\,\,
   M \Big[
\Eq(zzz)
$$
  we have:
$$
\eqalign{
&\P\big(T^{c,z,K(N)}_{l, N}(B) \geq 2\big) \cr
&{}  \leq  \delta |B| K(N) (2/\sqrt{2\pi})e^{-(2|c|z+z^2)N/2}
       + \Big(|B| K(N) (2/\sqrt{2\pi})e^{-(2|c|z+z^2)N/2}\Big)^{2}/2. }
\Eq(l3ii)
$$
}

\remark The bound \eqv(l3ii) is far from being
    the optimal one, but it is enough for our purpose.
  Therefore, we do not prove a precise bound
       that requires much more tedious computations.

\noindent{\it Proof}.
    The right-hand side of \eqv(l3i) is bounded from above by
$$ (\bar \a_{l+1}\cdots \bar \a_k)^N
        \P\Big(  |X - \sqrt{N}(|c|+z)|
      \in  K(N) e^{c^2 N/2} (\bar \a_{l+1} \cdots \bar \a_k)^{-N} B \Big)
\Eq(xxx)
$$
  with $X$ a standard Gaussian random variable.
   Since by the assumption of the lemma and by \eqv(cru)
   we have $ e^{c^2 /2} (\bar \a_{l+1} \cdots \bar \a_k)^{-1}<1$,
     then \eqv(l3i) is obvious from
 the trivial estimate \eqv(ko).

  To prove (ii), note  that
$\E T_{l, N}^{c,z, K(N)}(B)$ just equals
 \eqv(xxx), whence
$$
\E T_{l, N}^{c,z,K(N)}(B) \leq  (1+\delta) |B| K(N) (2/\sqrt{2\pi})e^{-(2|c|z+z^2)N/2}.
\Eq(yy)
$$
  Finally
$$  \P\big(T^{c,z,K(N)}_{l, N}(B) \geq 2\big) \leq \E T_{l, N}^{c,z,K(N)}(B)
     -  \Big(1- \P\big(T^{c,z,K(N)}_{l, N}(B)=0\big)\Big)
\Eq(xyx)
$$
        where  by Lemma \thv(lem2) $\P\big(T^{c,z,K(N)}_{l, N}(B)=0\big)$
      is bounded from above by the exponent \eqv(sh2n).
The assumption \eqv(zzz) and the fact that
   $|B|<1/K$ assure that the argument of this exponent
 is smaller than $1$ by  absolute value, i.e.
  $$
0<(1-\delta)|B| K(N) (2/\sqrt{2\pi})e^{-(2|c|z+z^2)N/2}<1-\delta.
\Eq(gggh)
$$
   Then \eqv(xyx), \eqv(yy), the  bound \eqv(sh2n)
 with \eqv(gggh) and the elementary fact
  that $e^{-x}\leq 1-x +x^2/2$ for $0<x<1$
   yield  the estimate \eqv(l3ii). \endproof

    We are now ready to  prove the tightness \eqv(2d)
of the family
  $\{\MM_N^l\}_{N=1}^{\infty}$ for $l=1,\ldots, k-1$.
    For a given $\epsilon>0$,  let us first fix $y$ large enough
  and $N_1(y)$  such that
$$
 \P(\bar B_N^l(y))< \epsilon/4 \ \ \forall N\geq N_1=N_1(y),
\Eq(uu1nn)
$$
   which is possible due to \eqv(bb). Now let us split the segment
  $A=[a, b]$ into $R$ disjoint segments $A_1,\ldots, A_R$ of size
  $(b-a)/R$, $R>1$.  Then
$$\eqalign{
 & \P(\{\MM_N^l(A)>R\}\cap B_N^l(y)) \leq \sum_{i=1}^R
      \P(\{\MM_N^l(A_i)\geq 2\} \cap B_N^l(y)) \cr
 &   \leq \sum_{i=1}^R \sum_{\hat \s \in \Sigma_{w_l N} }
    \P( C_N^l(A_i, \wh \s) \cap B_N^l(y, \wh \s)) \cr
&\ \ {} + \sum_{i=1}^R \sum_{\hat \tau, \hat \eta \in \Sigma_{w_l N},
       \hat \tau \ne \hat \eta}
      \P( D_N^l(A_i, \wh \tau) \cap
            D_N^l(A_i, \wh \eta)
 \cap B_N^l(y, \wh \tau) \cap B_N^l(y, \wh \eta))} \Eq(gg.1) $$
     where
$$\eqalign{ & C_N^l(A_i, \wh \s)=\Big\{
   \exists \check \eta, \check \tau \in \Sigma_{(1-w_l) N},
   \check \eta \ne \check \tau : \cr
&\ \ \ \   \Big| \check X^{J_l}_{\hat \s \check \s}-\sqrt{N}
   \big[\widetilde c_l+N^{-1}(\Gamma_{l, N}- U_{l, N}^{-1} (\wh X_{\hat
        \s}^{J_l}))\big]\Big| \in R_l(N)^{-1} A_i
 \hbox{ for }\check \s=\check \eta, \check \s=\check \tau \Big\}, }
$$
$$ D_N^l(A_i, \wh \s)
     =\Big\{ \exists \check \s \in \Sigma_{(1-w_l) N}:
    \Big| \check X^{J_l}_{\hat \s \check \s}-\sqrt{N}
   \big[\widetilde c_l+N^{-1}(\Gamma_{l, N}- U_{l, N}^{-1} (\wh X_{\hat
        \s}^{J_l}))\big]\Big| \in R_l(N)^{-1} A_i\Big\}, \Eq(ccc)$$
and
$$
B^l_N(y, \wh \s )=\Big\{  \forall j=1,\dots, l :
 2\Gamma_{j, N} - 2 N d_j - (\gamma_1+\cdots +\gamma_j)y
         <U_{j,N}^{-1}(\wh X^{J_j}_{\hat\s})
    <y(\gamma_1+\cdots +\gamma_j) \Big\}.
\Eq(bbbn)
$$
   Each term in the first sum of \eqv(gg.1)
   equals
$$
\eqalign{
& \E \big [\1_{\big\{ B_N^l(y, \hat \s)\big\} }
       \E\big(\1_{\big\{C_N^l(A_i, \hat \s)\big\}}\bigm|
           \wh X_{\hat \s}^{J_j},
         \forall_{j=1}^{l} \big ) \big]\cr
& =
  \E \big[\1_{\big\{ B_N^l(y, \hat \s)\big\} }
       \E\big(\1_{\big\{ T_{l, N}^{c, z(\hat \s), K(N)}(A_i)
 \geq 2 \big\}}\mid
           \wh X_{\hat \s}^{J_j},
         \forall_{j=1}^{l} \big )\big ]}
\Eq(cexx)
$$
  with the random variables
     $T^{c, z, K(N)}_{l, N}$ defined in Lemma \thv(lem3)
  and  with  parameters $c, K(N),z(\wh \s)$
 defined  by \eqv(cc) and \eqv(zz).
            Furthermore, on $B_N^l(y, \wh \s)$,
  the parameter  $z(\wh \s)$ satisfies the condition
      \eqv(zzz) with the constant
     $K=e^{\beta_l(\gamma_1+\cdots +\gamma_l)y}$ and
  $M= 2d_l(\bar a_{l+1}+\cdots +\bar a_k)^{-1/2}+\epsilon$
       with some small $\epsilon>0$.
   Therefore, if $|A_i|=(a-b)/R
 < e^{-\beta_l(\gamma_1+\cdots+\gamma_l)y}$,
 then the  assertion (ii)
   of Lemma~\thv(lem3) applies to the conditional expectation
    in \eqv(cexx). Next,
  each term of the second sum of \eqv(gg.1) equals
   $$\eqalign{
& \E \Big[\1_{\big\{ B_N^l(y, \hat \eta), B_N^l(y, \hat \tau)\big\}}
       \E\big(\1_{\big\{D_N^l(A_i, \hat \eta)\big\}}\bigm|
           \wh X_{\hat \eta }^{J_j},
         \forall_{j=1}^l \big)  \E\big (\1_{\big\{D_N^l(A_i, \hat
           \tau)\big\}
       }\bigm|
            \wh X_{\hat \tau }^{J_j},
         \forall_{j=1}^l \big) \Big]\cr
& = \E \Big[\1_{\big\{ B_N^l(y, \hat \eta), B_N^l(y, \hat \tau)\big\}}
       \E\big(\1_{\big\{T_{l, N}^{c, z(\hat \eta), K(N)}(A_i) \geq 1
         \big\}}
    \bigm|
           \wh X_{\hat \eta }^{J_j},
         \forall_{j=1}^l \big) \cr
 &\ \ \ \ \ {}\times \E\big(\1_{\big\{ T_{l,
 N}^{c, z(\hat \tau), K(N)}(A_i) \geq 1
\big\}}\bigm|
            \wh X_{\hat \tau }^{J_j},
         \forall_{j=1}^l \big) \Big] }
\Eq(xxc2)
$$
  where on   $B_N^l(y, \hat \eta) \cap B_N^l(y, \hat \tau)$
  we have
      $-\epsilon < z(\wh \tau), z(\wh \eta)<
      2 d_l(
 \bar a_{l+1}+\cdots + \bar a_k )^{-1/2}
      +\epsilon$ with some small $\epsilon>0$.
   Then the  assertion (i) of Lemma \thv(lem3) applies
   to the conditional expectations in \eqv(xxc2).
  Thus  by Lemma \thv(lem3),  for any $\delta>0$,
  there exists $N_2(y, \delta)$ such that
  for all $N\geq N_2$
$$
 \eqalign{& \sum\limits_{i=1}^R
 \P( \{\MM_N^0(A_i)\geq 2\} \cap B_N^l(y)  )\cr
 & \leq \sum\limits_{i=1}^R
      \delta (2/\sqrt{2\pi})K(N) (b-a)R^{-1}
     \E \Big( \sum_{\hat \s \in \Sigma_{w_l N}}
           \1_{\big\{ B_N^l(y, \hat \s)\big\}}  e^{ - \big(2|c| z(\hat \s )+
          z^2(\hat \s ) \big)N/2 } \Big)\cr
 &\ \ \ \ \ {}+  \sum\limits_{i=1}^R
          (4/2\pi) K(N)^2 (b-a)^2 R^{-2} \cr
&\ \ \ \ \ \ \ \ \ \   {}\times  \E \Big(
    \frac{1}{2} \sum_{\hat \s \in \Sigma_{w_l N}}
        \1_{\big\{ B_N^l(y, \hat \s)\big\}}
          e^{ - \big(2|c| z(\hat \s )+
          z^2(\hat \s )\big) N }  \cr
&\ \ \ \ \ \ \ \ \ \ \ \ \ \ \  {} +
      \sum_{\hat \tau, \hat \eta \in \Sigma_{w_l N}:
   \hat \tau \ne \hat \eta }
 \1_{\big\{ B_N^l(y, \hat \tau),  B_N^l(y, \hat \eta)
\big\}}
e^{-
 \big(2|c| z(\hat \tau )+
          z^2(\hat \tau) +
  2|c| z(\hat \eta )+
          z^2(\hat \eta ) \big) N /2
 } \Big) \cr
&=\delta (b-a) I_N(y)+R^{-1}(b-a)^2J_N(y)/2 }
$$
where
$$ I_N(y)=
 (2/\sqrt{2\pi}) K(N) \E \Big( \sum_{\hat \s \in \Sigma_{w_l N}}
           \1_{\big\{ B_N^l(y, \hat \s)\big\}}  e^{ - \big(2|c| z(\hat \s )+
          z^2(\hat \s ) \big)N/2 } \Big),$$
$$ J_N (y)= (4/(2\pi))
K(N)^2  \E \Big(\sum_{\hat \s \in \Sigma_{w_l N}
           }
\1_{\big\{ B_N^l(y, \hat \s)\big\}}
e^{ - \big(2|c| z(\hat \s )+
          z^2(\hat \s ) \big)N/2 }
\Big)^2.
$$
   Here, the quantity  $I_N(y)$  converges
   to
    $$\eqalign{
    I(y) & = \E \int\limits_{\forall 1\leq j\leq l: \atop
    \gamma_1x_1+\cdots +\gamma_j x_j <(\gamma_1+\cdots \gamma_j)y }
   e^{\beta (\gamma_1x_1+\cdots \gamma_l x_l)} \PP_l(dx_1\dots, dx_l)\cr
   & = \int\limits_{\forall 1\leq j\leq l:\atop
    \gamma_1x_1+\cdots +\gamma_j x_j <(\gamma_1+\cdots \gamma_j)y }
   e^{\beta (\gamma_1x_1+\cdots \gamma_l x_l)-x_1-\cdots-x_l}
   dx_1,\dots, dx_l<\infty.}$$
       Therefore,  one can fix $N_3=N_3(y)$ large enough
  and  then $\delta=\delta(y)$ so small that
 $ \delta(b-a)J_N^1(y)<\epsilon/4$,  $\forall N \geq N_3(y).$
   The term $J_N(y)$ converges to
$$
J(y)=\E \Big(\int\limits_{\forall j=1,\dots,l: \atop
   (\gamma_1 x_1+\cdots +\gamma_j x_j)<(\gamma_1+\cdots +\gamma_j)y }
     e^{\beta_l(\gamma_1x_1+\cdots +\gamma_lx_l)}\PP^l(dx_1\cdots
     dx_l)\Big)^2
\Eq(ab.rrr.1)
$$
      which is finite.  In fact, $J(y)$  is the sum of $l+1$ terms,
  the $k$th  of them being
$$
\eqalign{
&2^{\1_{ k\ne l }}\!\!\!\!\!\!\!\!\!\!\!\!\!\!\!\!\!
\int\limits_{{\forall 1\leq i\leq k :
  (\gamma_1x_1+\cdots \gamma_i x_i)<(\gamma_1+\cdots +\gamma_i)y  \atop
   \forall k+1 \leq i \leq l : (\gamma_1x_1+\cdots+\gamma_k x_k+\cdots
     \gamma_i v_i)<(\gamma_1+\cdots +\gamma_i)y }
\atop  \forall k+1 \leq i \leq l : (\gamma_1x_1+\cdots+\gamma_k x_k+\cdots
     \gamma_i w_i)<(\gamma_1+\cdots +\gamma_i)y
     }\!\!\!\!\!\!\!\!\!\!\!\!\!\!
\!\!\!\!\!\!\!
    e^{2 \beta_l(\gamma_1x_1+\cdots +\gamma_k x_k)}
    e^{ \b_l(\gamma_{k+1}v_{k+1}+\cdots +\gamma_l v_l)}
      e^{ \b_l(\gamma_{k+1}w_{k+1}+\cdots +\gamma_l w_l)}
\cr&
 \times  e^{-x_1-\cdots -x_k-v_{k+1}-\cdots -v_l-w_{k+1}-\cdots -w_l}
    dx_1\cdots dx_k dv_{k+1}\cdots dv_l dw_{k+1}\cdots dw_l<\infty.
}
\Eq(lm)
$$
    Then for any $\epsilon>0$, one can choose $N_4=N_4(y)$
  such that for all $N\geq N_4(y)$
  $|J_N(y) - J(y)|<\epsilon/4$.
       Next, let us choose $R_0=R_0(y)>K= e^{\beta(\gamma_1+\cdots
         \gamma_l)y}(b-a)$ such that
 $ (b-a)^2 R^{-1}_0<1$ and also such that
      $(b-a)^2 R^{-1}_0 J(y)<\epsilon/4$.
    Thus $(b-a)^2 R^{-1}J_N(y)/2<\epsilon/2$
   $\forall N \geq N_4(y)$ and  $\forall R \geq R_0$.
   Hence,
$$
 \sum\limits_{i=1}^R
 \P( \{\MM_N^0(A_i)\geq 2\} \cap B_N^l(y)  )< 3\epsilon/4
           \ \ \ \forall R\geq R_0, \hbox{ and }
   \forall N \geq  N_2(\delta(y), y), N_3(y), N_4(y).
\Eq(sjs)
$$
 Taking into account \eqv(uu1nn),  we obtain that
$$
 \P(\MM_N^l(A)>R) <\epsilon \ \ \ \forall R\geq R_0 \hbox{ and }
     \forall N \geq \max(N_1, N_2, N_3, N_4),
\Eq(xsx)
$$
 whence
$$
\P(\MM_N^l(A)>\max(R_0, 2^{N_1}, 2^{N_2}, 2^{N_3}, 2^{N_4}))
   <\epsilon \ \ \forall N\geq
1,
\Eq(ab.sjs.1)
$$
 then $\MM_N^l$ is tight.

      It remains to show that the limit $\tilde \MM^l$
   of any weakly
  convergent subsequence of $\{\MM_N^l\}$ is a simple process,
  that is very easy. Consider any segment
    $A=[a, b)$ and its dissecting system
   $\{A_{r,i},i=1,2,\dots, 2^r, r=1,2,\dots \}$ such that
    $A_{1,1}=[a, (a+b)/2)$ and  $A_{1,2}=[(a+b)/2, b)$
     are obtained by splitting $[a, b)$ in the middle
    and the system of disjoint intervals
    $\{A_{r,i}, i=1,2,\dots, 2^{r}\}$ is obtained
    from $\{A_{r-1,i}, i=1,2,\dots, 2^{r-1}\}$
   by splitting similarly each segment of the latter system
    into two parts in the middle. It follows from the estimates
     \eqv(uu1nn) and \eqv(sjs)   that for any $\epsilon>0$
    there exists $N_0$ and  $r_0$ such that
$$
\P(\exists i=1,\dots, 2^r: \MM_N^l(A_{r,i})\geq 2)<\epsilon
    \ \ \forall N\geq N_0, \ \forall r\geq r_0.
\Eq(ab.sjs.2)
$$
   Then for any $\epsilon>0$ there exists $r_0$ such that
$$
\P(\exists i=1,\dots, 2^r : \tilde \MM^l(A_{r,i})\geq 2)<\epsilon
    \ \ \ \forall r\geq r_0.
\Eq(ab.sjs.3)
$$
  Then $\tilde \MM^l$ can have double points
  within $A$ with probability smaller than $\epsilon$.
  Since $\epsilon>0$ is arbitrary,
     it follows that $\tilde \MM^l$ is simple.
 Thus the proof of the theorem is complete.
\endproof

\bigskip
{\headline={\ifodd\pageno\rightheadline \else \leftheadline \fi}}
\def\rightheadline{\it  {Mertens}\hfil\tenrm\folio}
\def\leftheadline{\tenrm \folio \hfil\it  {References}}

\Refs

\item{[BaMe]}
H.~Bauke and St. Mertens.
  Universality in the level statistics of disordered systems.
  {\it Phys. Rev. E}, 70:025102(R), 2004.

\item{[BK1]}
A.~Bovier and I.~Kurkova.
  Derrida's generalized random energy models. {I}. {M}odels with
  finitely many hierarchies.
  {\it Ann. Inst. H. Poincar\'e Probab. Statist.}, 40(4):439--480,
  2004.

\item{[BK2]}
A.~Bovier and I.~Kurkova.  Local energy statistics in disordered
system: a proof of the local REM conjecture. Preprint of the
  University Paris~6, April (2005).

\item{[BKL]} A.~Bovier, I.~Kurkova, M.~Lowe. Fluctuations of the free
  energy in the REM and the $p$-spin SK models, {\it Ann. Probab.}
   {\bf 30} (2002) 605--651.

\item{[DV]} D.J.~Daley, D.~Vere-Jones, {\it An introduction
     to the theory of point processes.} Springer Series in
 Statistics, Springer-Verlag (1988).

\item{[Der1]}
B.~Derrida.
 Random-energy model: an exactly solvable model of disordered systems.
 {\it Phys. Rev. B (3)}, 24(5):2613--2626, 1981.

\item{[Der2]}
B.~Derrida.
 A generalization of the random energy model that includes
  correlations between the energies.
 {\it J. Phys. Lett.}, 46:401--407, 1985.

\item{[LLR]}
M.R. Leadbetter, G.~Lindgren, and H.~Rootz{\'e}n.
{\it Extremes and related properties of random sequences and
  processes}.
 Springer Series in Statistics. Springer-Verlag, New York, 1983.

\item{[Ka]} O.~Kallenberg, Random Measures, fourth ed., Akademie
  Verlag, Berlin, 1986.

\endRefs 

\end